\documentclass[conference]{IEEEtran}
\usepackage{float}
\usepackage{placeins}
\IEEEoverridecommandlockouts

\usepackage{cite}
\usepackage{amsmath,amssymb,amsfonts}
\usepackage{algorithmic}
\usepackage{graphicx}
\usepackage{textcomp}
\usepackage{xcolor}
\usepackage{bm}
\usepackage{hyperref}
\usepackage{braket}
\providecommand{\outerproduct}[2]{|#1\rangle\langle #2|}
\def\BibTeX{{\rm B\kern-.05em{\sc i\kern-.025em b}\kern-.08em
T\kern-.1667em\lower.7ex\hbox{E}\kern-.125emX}}

\usepackage{placeins} 
\IEEEoverridecommandlockouts 

\begin{document}

\title{An Adaptive Purification Controller for Quantum Networks: Dynamic Protocol Selection and Multipartite Distillation \thanks{Sponsored in part by the Bavarian Ministry of Economic Affairs, Regional Development and Energy as part of the 6GQT project.}}

\author{
    \IEEEauthorblockN{1\textsuperscript{st} Pranav Kulkarni}
    \IEEEauthorblockA{\textit{AIGNOSCO GmbH} \\
    Munich, Germany \\
    pranav.kulkarni@aignosco.com}
    \and
    \IEEEauthorblockN{2\textsuperscript{nd} Leo Sünkel}
    \IEEEauthorblockA{\textit{LMU Munich} \\
    Munich, Germany \\
    leo.suenkel@ifi.lmu.de}
    \and
    \IEEEauthorblockN{3\textsuperscript{rd} Michael Kölle}
    \IEEEauthorblockA{\textit{AIGNOSCO GmbH} \\ \textit{LMU Munich} \\
    Munich, Germany \\
    michael.koelle@aignosco.com}
}

\maketitle


\begin{abstract}
    Efficient entanglement distribution is a cornerstone of the Quantum Internet. However, physical link parameters such as photon loss, memory coherence time, and gate error rates fluctuate dynamically, rendering static purification strategies suboptimal. In this paper, we propose an Adaptive Purification Controller (APC) that automatically optimizes the entanglement distillation sequence to maximize the goodput, i.e., the rate of delivered pairs meeting a strict fidelity threshold. By treating protocol selection as a resource allocation problem, the APC dynamically switches between purification depths and protocols (BBPSSW vs. DEJMPS) to navigate the trade-off between generation rate and state quality. Using a dynamic programming planner with Pareto pruning, simulation results show that our approach mitigates the "fidelity cliffs" inherent in static protocols and reduces resource wastage in high-noise regimes. Furthermore, we extend the controller to heterogeneous scenarios, and evaluate it for both multipartite GHZ state generation and continuous-variable systems using effective noiseless linear amplification models. We benchmark its computational overhead, showing decision latencies in the millisecond range per link in our implementation.
\end{abstract}


\section{Introduction}
Long-distance quantum communication and distributed quantum computing rely on the distribution of high-fidelity entanglement over noisy channels with finite coherence times. Physical channels and devices introduce loss, dephasing, and gate errors, so raw Bell pairs or multipartite entangled states must be transformed into fewer but more entangled copies using local operations and classical communication (LOCC)~\cite{Bennett1996Purification,Deutsch1996QPA}. In repeater architectures, entanglement generation, purification, and swapping are interleaved along multi-hop paths, and the timing of these operations interacts non-trivially with channel round-trip times, detector efficiencies, and memory decoherence.

Several quantum network simulators have emerged to study these effects, including NetSquid~\cite{NetSquid2021}, SimQN~\cite{SimQN2023}, SeQUeNCe~\cite{Sequence}, and QuNetSim~\cite{QuNetSim}. Many provide circuit-level or fidelity-level physical models for entanglement generation and swapping, and some include single-link purification protocols or link-layer purification selection modules. However, purification is often configured in a static way, such as a fixed number of BBPSSW rounds per link, or treated as a purely local primitive without a global optimization over routed paths or over different protocol families (BBPSSW versus DEJMPS versus higher-order recurrence or hashing). Recent work on purification protocol selection integrates device-level noise models and density matrix simulation in a link-layer module~\cite{shi2024designentanglementpurificationprotocol}, but does not couple this to a path-level planner that co-optimizes purification, swapping, and timing.

This paper introduces an Adaptive Purification Controller (APC), a module in the KOSMOS (Koordinations-, Optimierungs- und Simulations-Software f\"ur Modulare Quanten-Systeme in 6G Netzwerken) quantum network simulator that leverages the low latency expected of future 6G quantum technology. The APC sits between routing and the quantum execution driver. Given a routed path and a target end-to-end fidelity, the APC:
\begin{itemize}
  \item models each link using a fidelity-based Werner or Bell-diagonal description and closed-form recurrence primitives,
  \item enumerates per-hop recurrence depth and protocol choices using a dynamic programming planner with Pareto pruning,
  \item supports multiple objectives, including minimizing time, minimizing raw entanglement cost, Pareto trade-offs, and maximizing goodput,
  \item incorporates gate times, classical communication delays, Bell state measurement success probabilities, and memory decoherence,
  \item optionally applies multipartite GHZ stabilizer distillation as a post stage after bipartite arms are established,
  \item optionally applies continuous variable distillation using effective models of Gaussification and noiseless linear amplification.
\end{itemize}

The APC module exposes a narrow public interface through an \texttt{APCController} and configuration dataclasses. The planner and physics layers are separated so that new physical models can be added without changing the planning logic.

\textbf{Paper organization and contributions.}
Section~II reviews bipartite recurrence purification, entanglement swapping, and key noise models used in repeater chains.
Section~III summarizes related work on purification scheduling and adaptive repeater control.
Section~IV presents the Adaptive Purification Controller (APC) architecture and planning formulation, including the physics primitives, objective functions, and the path-level planner used in KOSMOS.
Section~V describes the experimental setup and simulation parameters.
Section~VI reports quantitative results that characterize (i) the fidelity--distance feasibility region, (ii) the ``noise cliff'' regime where additional purification becomes counterproductive, (iii) the impact of memory coherence time \(T_2\), (iv) protocol selection effects (BBPSSW vs.\ DEJMPS), (v) multipartite GHZ extensions, and (vi) planning latency.
Section~VII concludes and outlines future work.

Our main contributions are:
\begin{itemize}
  \item An \emph{Adaptive Purification Controller} module integrated into KOSMOS that computes per-path purification and swapping plans under a target fidelity constraint \(F^\star\).
  \item A fast planner that propagates and prunes a compact set of candidate partial plans (a Pareto frontier) across a path, enabling near-real-time decision-making in heterogeneous networks.
  \item A unified interface for bipartite (BBPSSW/DEJMPS) recurrence primitives and optional multipartite (GHZ) and continuous-variable (CV) post-processing models within the same planning framework.
  \item An empirical characterization of operating regimes where adaptive depth and protocol selection significantly improve delivered goodput over fixed-depth baselines.
\end{itemize}

\section{Background}
\label{sec:background}

\subsection{Entanglement Purification Protocols}
Entanglement purification aims to transform many copies of a noisy entangled state into fewer copies of a more entangled state using local operations and classical communication (LOCC). For bipartite qubit systems, two widely used families of recurrence protocols are:
\begin{itemize}
  \item the BBPSSW protocol of Bennett, Brassard, Popescu, Schumacher, Smolin and Wootters~\cite{Bennett1996Purification}, which assumes Werner or isotropic input states,
  \item the DEJMPS protocol of Deutsch, Ekert, Jozsa, Macchiavello, Popescu and Sanpera~\cite{Deutsch1996QPA}, which operates on Bell-diagonal states and can handle non-Werner noise more robustly.
\end{itemize}

A Werner state with target Bell state $\ket{\Phi^+}$ can be written as
\begin{equation}
  \rho_W(F) = F \outerproduct{\Phi^+}{\Phi^+}
  + \frac{1-F}{3} \sum_{k\in\{\Phi^-,\Psi^+,\Psi^-\}}
  \outerproduct{k}{k}
\end{equation}
where $F$ is the fidelity with respect to $\ket{\Phi^+}$. In the ideal BBPSSW protocol, two copies of $\rho_W(F)$ are processed by bilateral CNOT gates and measurements. One copy is kept if the measurement outcomes coincide at both ends. For Werner inputs, the fidelity and success probability update in closed form~\cite{Bennett1996Purification}:
\begin{align}
  p_{\text{succ}}^{\text{BBP}}(F)
  &= F^2 + \frac{2}{3} F (1-F) + \frac{5}{9}(1-F)^2,
  \label{eq:bbp-psucc}\\
  F_{\text{BBP}}'(F)
  &= \frac{F^2 + \frac{1}{9}(1-F)^2}{p_{\text{succ}}^{\text{BBP}}(F)}.
  \label{eq:bbp-fprime}
\end{align}
For $F>F_{\text{th}}$ with threshold $F_{\text{th}} \approx 1/2$, one has $F_{\text{BBP}}'(F) > F$, so iterating~\eqref{eq:bbp-psucc}--\eqref{eq:bbp-fprime} produces a sequence of fidelities converging toward unity.

DEJMPS operates on Bell-diagonal states of the form
\begin{equation}\label{eq:bell_diag}
  \rho_B(\bm{\lambda}) = \sum_{i=1}^4 \lambda_i \outerproduct{\Phi_i}{\Phi_i}, \qquad
  \sum_i \lambda_i = 1,
\end{equation}

For a single DEJMPS round, the coefficients update according to
\begin{align}\label{eq:dejmps_update}
  p_{\text{succ}}^{\text{DEJ}}(\bm{\lambda})
  &= (\lambda_1+\lambda_4)^2 + (\lambda_2+\lambda_3)^2, \\
  \lambda_1' &= \frac{\lambda_1^2 + \lambda_4^2}{p_{\text{succ}}^{\text{DEJ}}},\quad
  \lambda_2' = \frac{\lambda_2^2 + \lambda_3^2}{p_{\text{succ}}^{\text{DEJ}}},\\
  \lambda_3' &= \frac{2 \lambda_2 \lambda_3}{p_{\text{succ}}^{\text{DEJ}}},\quad
  \lambda_4' = \frac{2 \lambda_1 \lambda_4}{p_{\text{succ}}^{\text{DEJ}}}.
\end{align}

\textbf{Bell-basis indexing and implementation mapping.} To remove ambiguity in the coefficients $\lambda_i$ in~\eqref{eq:bell_diag}--\eqref{eq:dejmps_update},
we fix the Bell basis as
\begin{align}
\ket{\Phi^{\pm}} &= (\ket{00}\pm\ket{11})/\sqrt{2}, &
\ket{\Psi^{\pm}} &= (\ket{01}\pm\ket{10})/\sqrt{2},
\end{align}
and we take $(\ket{\Phi_1},\ket{\Phi_2},\ket{\Phi_3},\ket{\Phi_4})
=(\ket{\Phi^{+}},\ket{\Psi^{+}},\ket{\Psi^{-}},\ket{\Phi^{-}})$ so that
$\lambda_1=\bra{\Phi^{+}}\rho_B\ket{\Phi^{+}}$ and the Werner special case corresponds to
$\lambda_1=F$ and $\lambda_{2,3,4}=(1-F)/3$.

In the KOSMOS implementation, Bell-diagonal states are stored as a tuple
$\mathrm{BellDiagonal}(a,b,c,d)$ ordered as $(\Phi^{+},\Psi^{+},\Psi^{-},\Phi^{-})$. Therefore, when relating \eqref{eq:bell_diag}--\eqref{eq:dejmps_update} to code, the mapping is $(a,b,c,d)=(\lambda_1,\lambda_2,\lambda_3,\lambda_4)$.

\textbf{Implementation note (local basis choice).} The DEJMPS primitive used by APC assumes this Bell-state ordering (equivalently, a fixed choice of local basis). It does not apply any additional coefficient sorting/permutation step before evaluating~\eqref{eq:dejmps_update}.

Let $F^{(0)}$ denote the input fidelity and $F^{(j)}$ the fidelity after $j$ recurrence rounds (for either BBPSSW or DEJMPS) with corresponding per round success probabilities $p_{\text{succ}}^{(j)}$. If $r$ rounds are applied, the overall success probability and the expected raw pair consumption per surviving pair are
\begin{align}
  P_{\text{succ}}^{(r)} &= \prod_{j=0}^{r-1} p_{\text{succ}}^{(j)}, \label{eq:succ-multi-round}\\
  C_{\text{pairs}}^{(r)} &= \frac{2^{r}}{P_{\text{succ}}^{(r)}},
  \label{eq:pair-cost}
\end{align}
which we use as a definition of entanglement cost in APC.

Beyond recurrence, hashing-style one-way protocols achieve asymptotic distillation rates determined by entanglement measures such as coherent information~\cite{DevetakWinter2005}. APC currently focuses on finite-depth recurrence with small $r$ and treats hashing as future work.

For multipartite systems, D\"ur, Aschauer and Briegel developed entanglement purification protocols for two-colorable graph states, including GHZ and cluster states~\cite{Dur2003GraphStates, Aschauer2005TwoColorable}. These protocols use local Clifford operations and bilateral CNOTs to measure stabilizer parities and discard states with wrong syndromes. de Bone \emph{et al.} proposed protocols that create and distill GHZ states from Bell pairs via non-local stabilizer measurements using ancilla circuits~\cite{deBone2020GHZ}. APC implements an effective fidelity update that approximates these stabilizer-based protocols but delegates their detailed circuit structure to future work.

\subsection{Continuous Variable Entanglement Distillation}
Continuous variable (CV) entanglement between two modes can be modeled by two-mode squeezed states
\begin{equation}
  \ket{\psi(r)} = \sqrt{1-\lambda^2} \sum_{n=0}^{\infty} \lambda^n \ket{n,n}, \qquad \lambda = \tanh r,
\end{equation}
where $r$ is the squeezing parameter. Loss and Gaussian noise degrade CV entanglement, typically quantified by measures such as logarithmic negativity. Protocols for its distillation include Gaussification schemes and non-Gaussian operations that use photon subtraction or conditional measurements~\cite{Eisert2004CV, Fiurasek2009Gaussian}.

Heralded noiseless linear amplification (NLA) implements a non-deterministic gain operation $g>1$ that approximately maps coherent states as $\ket{\alpha} \mapsto \ket{g\alpha}$ while suppressing added noise~\cite{Xiang2010NLA}. In idealized models, a sequence of $K$ NLA stages yields a net gain $g^K$ with success probabilities that often scale as
\begin{equation}
  p_{\text{succ}}^{\text{NLA}} \propto \left(\frac{p_0}{g^2}\right)^K,
\end{equation}
where $p_0$ depends on beam splitter parameters and detector efficiencies. Modern work has derived tight upper bounds on the success probability of optimal NLA devices and constructed linear optics schemes that asymptotically saturate these bounds~\cite{Guanzon2024NLA}. APC uses these ideas to construct an effective CV distillation model that captures the main dependence of fidelity and success probability on $r$, $g$, and $K$ without simulating full Wigner functions.

\subsection{Entanglement Swapping and Fidelity-based Models}
Entanglement swapping transforms entangled pairs $(A,B)$ and $(C,D)$ into an entangled pair $(A,D)$ via a Bell state measurement on $(B,C)$~\cite{Zukowski1993Swapping}. Fidelity-based entanglement models used in high-level simulators such as SimQN\cite{SimQN2023} represent each link by a Werner parameter $w$ related to fidelity by
\begin{equation}
  w = \frac{4F - 1}{3}, \qquad F = \frac{1+3w}{4}.
  \label{eq:werner-conv}
\end{equation}
Swapping two Werner states with parameters $w_1,w_2$ yields another Werner state with parameter $w' = w_1 w_2$, so that
\begin{equation}
  F' = \frac{1+3 w_1 w_2}{4}.
  \label{eq:swap-fidelity}
\end{equation}
This multiplicative rule is a standard approximation in fidelity-based network models~\cite{SimQN2023}.

For probabilistic entanglement generation, many works assume that each generation attempt on link $i$ succeeds with probability $p_i$ and takes time $t_{0,i}$, so the number of attempts $X_i$ before success is geometric with mean $\mathbb{E}[X_i] = 1/p_i$. If $H$ links attempt generation in parallel, the time until all succeed is proportional to the maximum of $H$ geometric random variables. Its expectation can be expressed as~\cite{Shchukin2017Waiting}
\begin{equation}
  \mathbb{E}[\max(X_1,\dots,X_H)]
  = \sum_{k=1}^{\infty} \left[ 1 - \prod_{i=1}^H \left(1-(1-p_i)^{k-1}\right) \right],
  \label{eq:max-geometric}
\end{equation}
and can be evaluated numerically. Markov chain approaches further refine this picture when probabilistic swaps and finite memories are included~\cite{Shchukin2017Waiting, 2023Purification}.

APC adopts~\eqref{eq:swap-fidelity} and~\eqref{eq:max-geometric} as the backbone of its fidelity and timing models at the path level, with configurable shortcuts that approximate~\eqref{eq:max-geometric} by either a geometric scaling $1/p_{\min}$ or a max over identical link statistics in fair toy settings.

\section{Related Work}
\label{sec:related}
NetSquid~\cite{NetSquid2021} is a discrete-event simulator for quantum networks that supports state vector and density matrix simulations, detailed device noise, and memory decoherence. It allows implementation of entanglement distillation at the circuit level using primitive gates and measurements, and recent modules built on NetSquid target fidelity-aware repeater chain and entanglement distribution switch simulations, with configurable swapping and distillation mechanisms~\cite{NetSquidFidelity2024, NetSquidEBN2025}. These works provide powerful building blocks for studying routing and topology-level trade-offs but leave the choice of how many distillation rounds to perform on each link, and where, to user-supplied protocol logic.

SimQN~\cite{SimQN2023} offers both state-based and fidelity-based physical models, including an entanglement model with fidelity tracking and analytic operations for entanglement generation, swapping, and simple distillation. Its entanglement model provides closed-form recurrence maps for Bennett-style distillation on Werner and mixed Bell diagonal states, exponential decoherence of the Werner parameter in memory and over channels, and analytic swapping rules. The model is designed for speed and scalability, but it does not include a built-in dynamic planner for per-hop protocol choice or multipartite GHZ and CV distillation.

SeQUeNCe~\cite{Sequence} is an event-driven quantum network simulator that includes an entanglement management module with BBPSSW-style purification and entanglement swapping. Purification is available as a protocol component (\texttt{BBPSSWProtocol} and \texttt{BBPSSW\_BDS}), and tutorials demonstrate fixed round schemes on short repeater chains. The Bell diagonal variant can be configured to behave as BBPSSW (with twirling) or DEJMPS (without twirling). \textit{Twirling applies random local Pauli rotations to symmetrize the state, effectively erasing specific error biases to yield a standard Werner state characterized solely by fidelity.} SeQUeNCe exposes a method that computes the updated fidelity for a given input fidelity based on these configurations. However, SeQUeNCe does not provide a global path-level planner that jointly optimizes per-hop method, number of rounds, and end-to-end latency or resource cost.

QuNetSim~\cite{QuNetSim} focuses on high-level protocol logic and routing for quantum networks. While it supports entanglement distribution and routing, purification models are left to the user or delegated to backends, and no integrated purification planner is provided.

Beyond simulators, recent work on network-level purification and scheduling has studied how depolarizing noise, gate errors, storage times, and routing policies affect performance on complex topologies~\cite{2023Purification}. These works typically consider fixed purification protocols and focus on where to purify and how often. A complementary line of work optimizes repeater chain policies or cutoff thresholds using Markov decision processes or dynamic programming, but often without explicit purification choices. Y. Shi \emph{et al.}~\cite{shi2024designentanglementpurificationprotocol} proposed a purification protocol selection module for bipartite links that uses density matrix simulations under realistic errors to build heuristics for choosing BBPSSW, DEJMPS, or other protocols at the link layer. Their module optimizes over a local configuration space per hardware instance but does not combine these choices into a path-level planner that accounts for swapping and end-to-end objectives.

In this landscape, APC aims to bridge the gap between physics-accurate purification models and network-aware planning. It encapsulates closed-form bipartite recurrence, entanglement swapping, multipartite GHZ distillation, and CV distillation in a reusable physics backend, and exposes them to a planner responsible for optimizing time, resource cost, and goodput on top of routed paths.

\section{Methodology}
\label{sec:methodology}

\subsection{System Overview}
The APC consists of the following main components:
\begin{itemize}
  \item \textbf{APCController}: Public interface providing a \texttt{plan} method. It accepts a routed path, a request specifying source, destination, and target fidelity, configuration parameters, and optional constraints, and returns a purification and swapping plan.
  \item \textbf{Planner}: A dynamic programming planner with Pareto pruning that explores per-hop recurrence choices and composes them with entanglement swapping.
  \item \textbf{Physics backend}: A fidelity-based physical model implementing recurrence updates, entanglement swapping, GHZ stabilizer distillation, and CV distillation.
  \item \textbf{Policy module}: A heuristic module that chooses DEJMPS versus BBPSSW and shallow versus deeper recurrence per-hop based on link and device parameters.
  \item \textbf{Primitives}: Closed-form (or effective) update maps for BBPSSW and DEJMPS, Werner parameter conversion, depolarizing noise, GHZ stabilizer passes, and CV Gaussification/NLA stages.
\end{itemize}

Routing is performed by KOSMOS' routing module. APC takes as input the selected path, per-link parameters (e.g., length, initial fidelity, Bell state measurement success probabilities, and classical delays), and returns a plan consisting of per-link recurrence choices and swaps with predicted end-to-end fidelity, success probability, makespan, and entanglement cost.

\subsection{Bipartite Recurrence and Noise Model}

\subsubsection{Static Recurrence}
On each link $\ell$, APC represents the current pair by either a scalar fidelity $F$ and Werner parameter $w$ or by a Bell-diagonal vector $\bm{\lambda}$, depending on the configured physical model. Conversion between fidelity and Werner parameter is given by Eq. \ref{eq:werner-conv}.

For each supported recurrence protocol $P\in\{\text{BBPSSW},\text{DEJMPS}\}$, the physics backend implements ideal maps
\begin{equation}
  (F,p_{\text{succ}}) \mapsto (F_P'(F),p_{\text{succ},P}(F))
\end{equation}
for Werner inputs using~\eqref{eq:bbp-psucc}--\eqref{eq:bbp-fprime}, and, where configured, Bell-diagonal updates $(\bm{\lambda},p_{\text{succ}}) \mapsto (\bm{\lambda}',p_{\text{succ}}')$ using the DEJMPS relations. For $r$ rounds on link $\ell$ under protocol $P$, APC computes the sequence $F^{(j+1)} = F_P'(F^{(j)})$ and the multi-round success probability $P_{\text{succ}}^{(r)}$ from~\eqref{eq:succ-multi-round}. The expected raw entanglement cost per surviving pair after $r$ rounds is $C_{\text{pairs}}^{(r)}$ in~\eqref{eq:pair-cost}.

\subsubsection{Local Operations and Memory Noise}
Local gate and measurement errors are modeled as Pauli-twirled depolarizing channels acting on the kept pair after each round. A two-qubit depolarizing channel is represented as
\begin{equation}
  \mathcal{E}_2(\rho) = (1-\lambda_{2q})\rho + \lambda_{2q}\frac{\mathbb{I}_4}{4},
\end{equation}
where $\lambda_{2q}\in[0,1]$ is the depolarizing mixing strength. In KOSMOS, the user-facing parameters $p_1$, $p_2$, and $p_{\mathrm{meas}}$ denote \emph{Pauli-twirled} per-operation error probabilities (single-qubit, two-qubit, and local measurement bit-flip, respectively), i.e., with total probability $p$ a uniformly random non-identity Pauli is applied (on the corresponding $n$-qubit subsystem) and with probability $1-p$ the identity is applied.

The backend converts these primitive Pauli error rates into an \emph{equivalent depolarizing} strength for
\begin{equation}
  \Delta_{\lambda}(\rho)=(1-\lambda)\rho+\lambda \frac{\mathbb{I}_d}{d}, \qquad d=2^n,
\end{equation}
using the standard mapping
\begin{equation}
  \lambda = \frac{d^2}{d^2-1}\,p = \frac{4^n}{4^n-1}\,p.
  \label{eq:pauli_to_depol_general}
\end{equation}
For the 1-qubit and 2-qubit cases used throughout APC this yields
\begin{equation}
\lambda_{1q}= \frac{4}{3}p_1,\qquad
\lambda_{2q}= \frac{16}{15}p_2,\qquad
\lambda_{\mathrm{meas}}\approx \frac{16}{15}\frac{p_{\mathrm{meas}}}{2},
\label{eq:pauli_to_depol}
\end{equation}
where the measurement term folds a local bit-flip probability into an effective two-qubit depolarizing kick (at leading order, two local readouts contribute to a joint check/BSM).

Independent depolarizing channels are composed by multiplying reliabilities,
$1-\lambda_{\mathrm{tot}}=\prod_i(1-\lambda_i)^{n_i}$. Under this conversion, $\lambda$ reaches 1 at $p\le (4^n-1)/4^n$; for larger p the formal equivalence would yield $\lambda>1$, so we clamp $\lambda\in[0,1]$ as a numerical guardrail (the regimes of interest satisfy $p\ll 1$ for realistic quantum repeaters).”

Assuming $n_{1q}$ single-qubit gates, $n_{2q}$ two-qubit gates, and $n_{\mathrm{meas}}$ measurements per recurrence
round, the per-round reliability factor is
\begin{align}
  r_{\text{round}}
  &= r_1^{n_{1q}} r_2^{n_{2q}} r_m^{n_{\mathrm{meas}}},\\
  r_1 &= 1-\lambda_{1q},\qquad
  r_2 = 1-\lambda_{2q},\qquad
  r_m = 1-\lambda_{\mathrm{meas}},
  \label{eq:r_round}
\end{align}
where $\lambda_{1q},\lambda_{2q},\lambda_{\mathrm{meas}}\in[0,1]$ (and hence $r_1,r_2,r_m\in[0,1]$).

Equivalently, the round induces an \emph{effective} two-qubit depolarizing strength
$\lambda_{\mathrm{round}}=1-r_{\mathrm{round}}$.
If a waiting-time depolarizing penalty
$\lambda_{\mathrm{wait}}$ is enabled, we combine them as
$\lambda_{\mathrm{tot}}=1-(1-\lambda_{\mathrm{round}})(1-\lambda_{\mathrm{wait}})$.

The backend composes ideal recurrence and local noise as follows:
\begin{enumerate}
  \item compute the ideal updated state ($F'$ or $\bm{\lambda}'$) and success probability $p_{\text{succ}}$,
  \item compute $\lambda_{\mathrm{tot}}$ from $(p_1,p_2,p_{\mathrm{meas}})$ and the per-round gate counts,
  \item apply $\Delta_{\lambda_{\mathrm{tot}}}$ to the kept pair:
  \begin{itemize}
    \item \textbf{Werner tracking:} apply the effective two-qubit depolarizing channel directly in the fidelity domain,
    \begin{equation}
      F' \;\leftarrow\; (1-\lambda_{\mathrm{tot}})\,F' + \lambda_{\mathrm{tot}}\frac{1}{4},
      \label{eq:werner_depol_fidelity_update}
    \end{equation}
    and, when needed, convert to/from the Werner parameter $w$ using Eq.~\ref{eq:werner-conv},
    \item \textbf{Bell-diagonal tracking:} $\bm{\lambda}'\leftarrow (1-\lambda_{\mathrm{tot}})\bm{\lambda}'+\lambda_{\mathrm{tot}}(\tfrac14,\tfrac14,\tfrac14,\tfrac14)$.
  \end{itemize}
\end{enumerate}

Memory decoherence is modeled as exponential relaxation with an effective coherence time $T_2^{\text{eff}}$:
\begin{equation}
  F(t) = \frac{1}{4} + \left( F(0) - \frac{1}{4} \right) e^{-t/T_2^{\text{eff}}}.
  \label{eq:decoherence}
\end{equation}
APC applies this relaxation over the effective dwell time of the kept pair during each round and during swap operations (local gate time plus any classical wait), as provided by the planner time model.

Per-round local operation time is modeled as
\begin{equation}
  t_{\text{round}} = n_{1q} t_{1q} + n_{\text{CNOT}} t_{\text{CNOT}} + n_{\text{meas}} t_{\text{meas}} + t_{\text{classical}},
\end{equation}
where $t_{\text{classical}}$ captures per-round classical messages required to compare measurement outcomes. Swap operations are modeled similarly with their own gate counts and classical delays.

\subsection{Multipartite GHZ Stabilizer Distillation}
APC includes a GHZ post-processing stage that can be invoked after bipartite arms from a hub to multiple leaf nodes have been purified. In the current implementation, we use a \emph{planner-grade} stabilizer-pass acceptance model inspired by stabilizer-measurement-based GHZ distillation for graph states~\cite{Dur2003GraphStates,Aschauer2005TwoColorable,deBone2020GHZ}. This model is designed to be conservative and computationally cheap rather than a circuit-exact simulation.

A noisy $N$-qubit GHZ state is modeled using an isotropic GHZ-diagonal approximation:
\begin{equation}
  \rho_{\mathrm{GHZ}} = F_{\mathrm{GHZ}} \outerproduct{\mathrm{GHZ}_N}{\mathrm{GHZ}_N}
  + (1-F_{\mathrm{GHZ}})\,\sigma_{\perp},
  \label{eq:ghz_iso}
\end{equation}
where $\ket{\mathrm{GHZ}_N} = (\ket{0}^{\otimes N} + \ket{1}^{\otimes N})/\sqrt{2}$ and
$\sigma_{\perp} = \frac{\mathbb{I} - \outerproduct{\mathrm{GHZ}_N}{\mathrm{GHZ}_N}}{2^N - 1}$
is the maximally mixed state on the orthogonal subspace. This model implicitly assumes that random Pauli twirling is applied between distillation rounds, yielding a GHZ-diagonal (stabilizer-eigenvalue) description compatible with stabilizer-based checks~\cite{Dur2003GraphStates,Aschauer2005TwoColorable,deBone2020GHZ}.

While the stabilizer group is generated by $Z_i Z_N$ and $X^{\otimes N}$, our implementation models one \emph{full stabilizer pass} per round consisting of $(N-1)$ weight-2 $Z_i Z_N$ checks and one global $X^{\otimes N}$ check, i.e., $m=N$ checks per pass. We account for the resource overhead by charging approximately one ancilla resource per check, i.e., $m$ ancilla resources per GHZ pass.

To enable millisecond-scale planning, the backend uses an aggregate acceptance model rather than an explicit circuit simulation. Let $F \equiv F_{\mathrm{GHZ}}$ be the current GHZ fidelity, and define the (clamped) effective per-check imperfection
\begin{equation}
  \varepsilon \;\triangleq\; \mathrm{clip}_{[0,1]}\!\Big((1-F_{\mathrm{anc}}) + p_{\mathrm{meas}}^{\mathrm{GHZ}}\Big),
\end{equation}
which aggregates ancilla infidelity and measurement bit-flip error into a single conservative parameter.

We approximate the probability that the \emph{ideal} (target) stabilizer-eigenvalue string passes all $m$ checks as
\begin{equation}
  p_{\mathrm{keep,good}} \;\approx\; (1-\varepsilon)^m.
\end{equation}
For the \emph{noise} component, we assume GHZ-diagonal weight is uniformly distributed over the $2^m-1$ non-target stabilizer-eigenvalue strings, and each check outcome is flipped independently with probability $\varepsilon$. Under this uniform-noise approximation, the average false-accept probability is
\begin{equation}
  p_{\mathrm{keep,bad}} \;\approx\; \frac{1 - (1-\varepsilon)^m}{2^m - 1},
\end{equation}
which is continuous in $\varepsilon$ and vanishes as $\varepsilon \to 0$.

The resulting (effective) block success probability and posterior fidelity are
\begin{align}
  p_{\mathrm{succ}}^{\mathrm{GHZ}} &\approx F\,p_{\mathrm{keep,good}} + (1-F)\,p_{\mathrm{keep,bad}},\\
  F_{\mathrm{GHZ,out}} &\approx \frac{F\,p_{\mathrm{keep,good}}}{p_{\mathrm{succ}}^{\mathrm{GHZ}}}.
\end{align}
This formulation allows APC to budget both latency (roughly $m$ LOCC check outcomes per pass) and ancilla overhead without tracking the full $2^N$-dimensional density matrix.

\subsection{Continuous Variable Distillation}
For continuous variable entanglement, APC implements an effective model for applying an NLA-based distillation stage to a two-mode squeezed state subjected to symmetric loss. The backend tracks an effective squeezing parameter $r$ (or equivalently $\lambda=\tanh r$) and a loss parameter $\eta$, and maps these to a fidelity proxy $F_{\text{CV}}$ with respect to an ideal entangled Gaussian resource via a monotonically increasing function of $r$ and $\eta$~\cite{Eisert2004CV}. A sequence of $K$ NLA stages with gain $g$ yields an updated effective squeezing $r'_{\text{eff}}$ and a success probability of the form
\begin{equation}
  p_{\text{succ}}^{\text{CV}} \approx A \left(\frac{1}{g^2}\right)^K,
\end{equation}
with prefactor $A$ tuned so that $p_{\text{succ}}^{\text{CV}}$ remains below known optimal bounds for the chosen gain and loss~\cite{Guanzon2024NLA}. The backend then returns
\begin{equation}
  F_{\text{CV,out}}, \quad p_{\text{succ}}^{\text{CV}}, \quad
  C_{\text{CV}}, \quad t_{\text{CV}},
\end{equation}
where $C_{\text{CV}}$ is an effective resource count and $t_{\text{CV}}$ a time cost, both modeled analogously to the qubit case. These values are compatible with the planner and can participate in trade-off decisions.

\subsection{Planner and Objectives}

\subsubsection{Per-Link Design Space and Pareto Frontiers}
For each link $\ell$ in a routed path, APC considers recurrence depths $r\in\{0,1,\dots,R_{\max}\}$, where $R_{\max}$ is a configurable maximum per-hop round count. A policy module chooses a method tag for each link and depth,
\begin{equation}
  u = (\ell,r,P) \in \mathcal{U}_\ell,
\end{equation}
where $P\in\{\text{BBPSSW},\text{DEJMPS}\}$ and optional tags such as \texttt{shallow} can be encoded by restricting $r$. For each candidate $u\in\mathcal{U}_\ell$, the physics backend provides the updated fidelity, success probability, entanglement cost, and time:
\begin{equation}
  \phi_\ell(u) = \big(F_{\text{out}}(\ell,u),\ p_{\text{succ}}(\ell,u),\ C_{\text{pairs}}(\ell,u),\ t(\ell,u)\big).
\end{equation}

APC constructs a per-link Pareto frontier $\mathcal{P}_\ell$ over these tuples with respect to two primary metrics: entanglement cost and time, under the constraint that $F_{\text{out}}(\ell,u)\geq F_{\min}$ for a configurable local threshold. A candidate $u$ is discarded if there exists another $u'$ with
\begin{equation}
  C_{\text{pairs}}(\ell,u') \leq C_{\text{pairs}}(\ell,u), \quad
  t(\ell,u') \leq t(\ell,u)
\end{equation}
and at least one strict inequality, while achieving comparable or higher fidelity.

\begin{figure*}[t!]
  \centering
  \includegraphics[width=\textwidth]{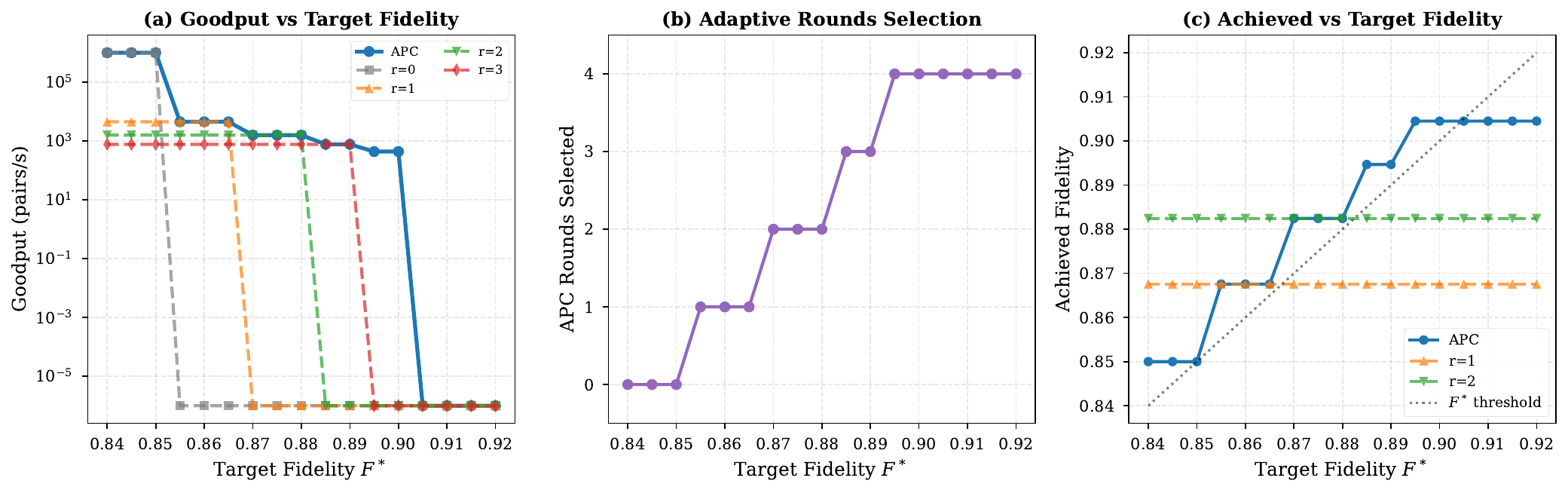}
  \caption{\textbf{Goodput optimization with APC on a single hop.} Setup: 15 km link, $F_0=0.85$, $T_2=100$ ms, target $F^\star \in [0.84,\,0.92]$. \textbf{(a)} Goodput (pairs per second, log scale) versus $F^\star$ for APC and fixed rounds $r=0,1,2,3$; APC tracks the upper envelope and declares infeasible beyond $F^\star \approx 0.885$. \textbf{(b)} Rounds selected by APC as a function of $F^\star$ (stepwise 0 to 4). \textbf{(c)} Achieved end-to-end fidelity versus target (45$^\circ$ threshold shown); APC meets or slightly exceeds $F^\star$ when feasible while fixed strategies under- or over-shoot.}
  \label{fig:goodput}
\end{figure*}

\begin{figure*}[t!]
  \centering
  \includegraphics[width=\textwidth]{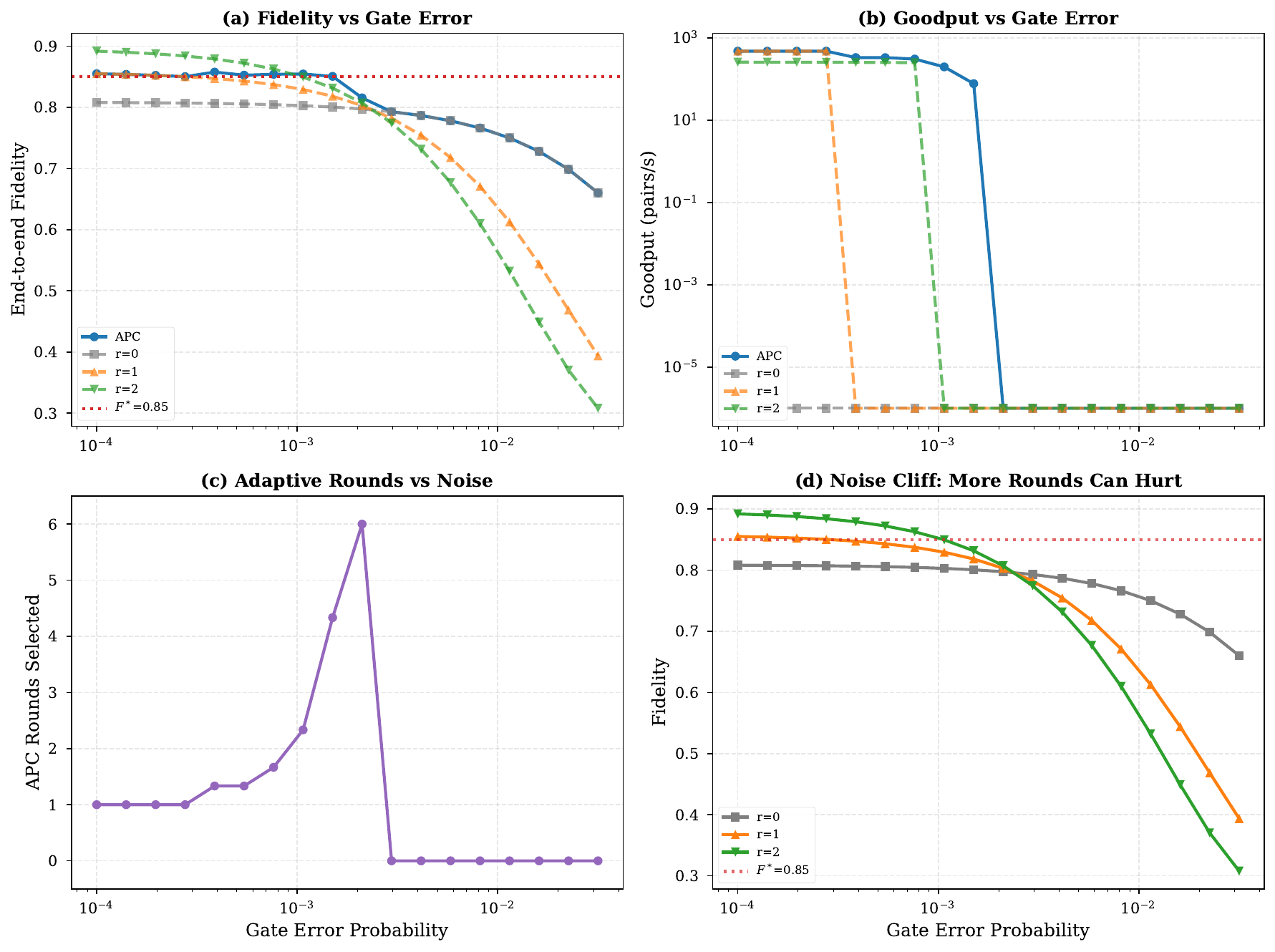}
  \caption{\textbf{APC rounds adapt to gate noise on a 3-hop chain.} Setup: total 24 km (3 links), $F_0=0.93$ per link giving $F_{\mathrm{raw}}\approx0.80$ after two swaps, target $F^\star=0.85$, gate or measurement error probability $\varepsilon$ swept from $10^{-4}$ to $3\times10^{-2}$. \textbf{(a)} End-to-end fidelity versus $\varepsilon$ with threshold $F^\star$ marked (to reduce parameter space, we use a correlated noise sweep $p_1=p_2=p_{\mathrm{meas}}=\varepsilon$). \textbf{(b)} Goodput versus $\varepsilon$ (log-log); APC sustains $\sim10^2$ to $5\times10^2$/s until the noise cliff. \textbf{(c)} APC-selected rounds (1 to 4) versus $\varepsilon$, compensating for increasing noise until saturation. \textbf{(d)} At high noise beyond the cliff, additional rounds reduce fidelity because CNOT-induced errors dominate; $r=0$ becomes least bad.}
  \label{fig:noise}
\end{figure*}

\begin{figure*}[t!]
  \centering
  \includegraphics[width=\textwidth]{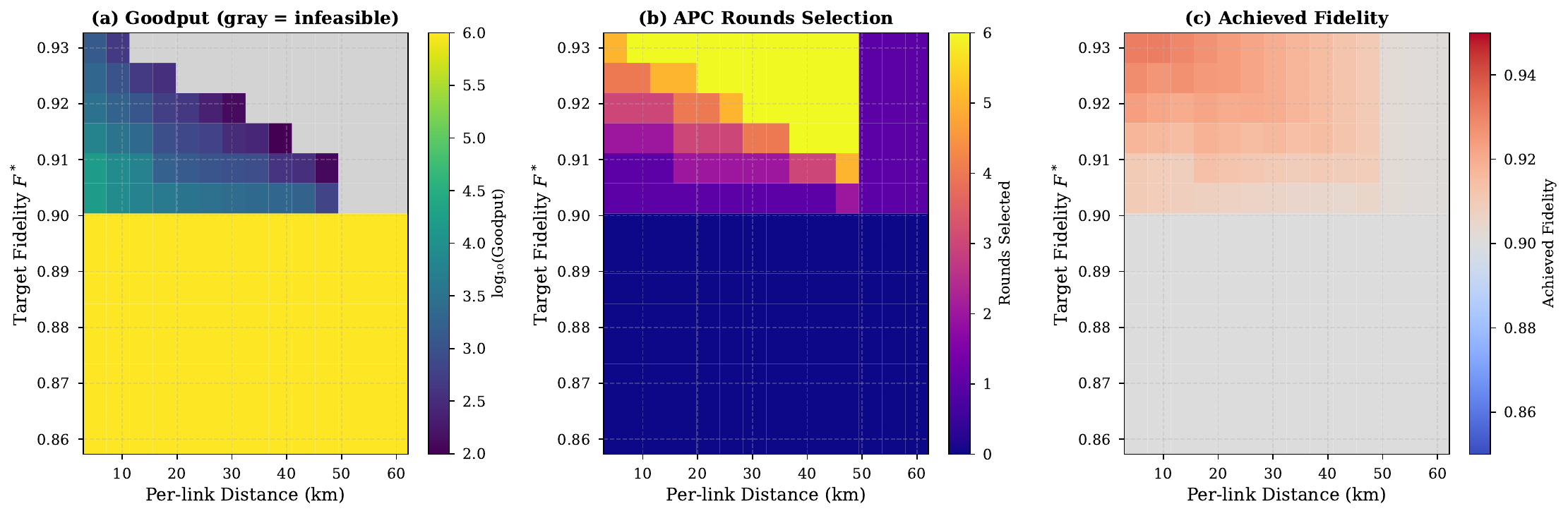}
  \caption{\textbf{Operating regime over distance and target fidelity.} Setup: 1 hop, $F_0=0.90$, $T_2=80$ ms; distance $5$ to $60$ km and target $F^\star$ in $[0.86,\,0.93]$. \textbf{(a)} Heatmap of $\log_{10}(\text{goodput})$ with gray for infeasible points. \textbf{(b)} APC-selected rounds $r\in[0,6]$ (banded regions show transitions $r=0\to1\to2\ldots$). \textbf{(c)} Achieved fidelity across the same grid; feasible cells meet $F^\star$, with gray indicating failure.}
  \label{fig:feasible}
\end{figure*}

\begin{figure*}[t!]
  \centering
  \includegraphics[width=\textwidth]{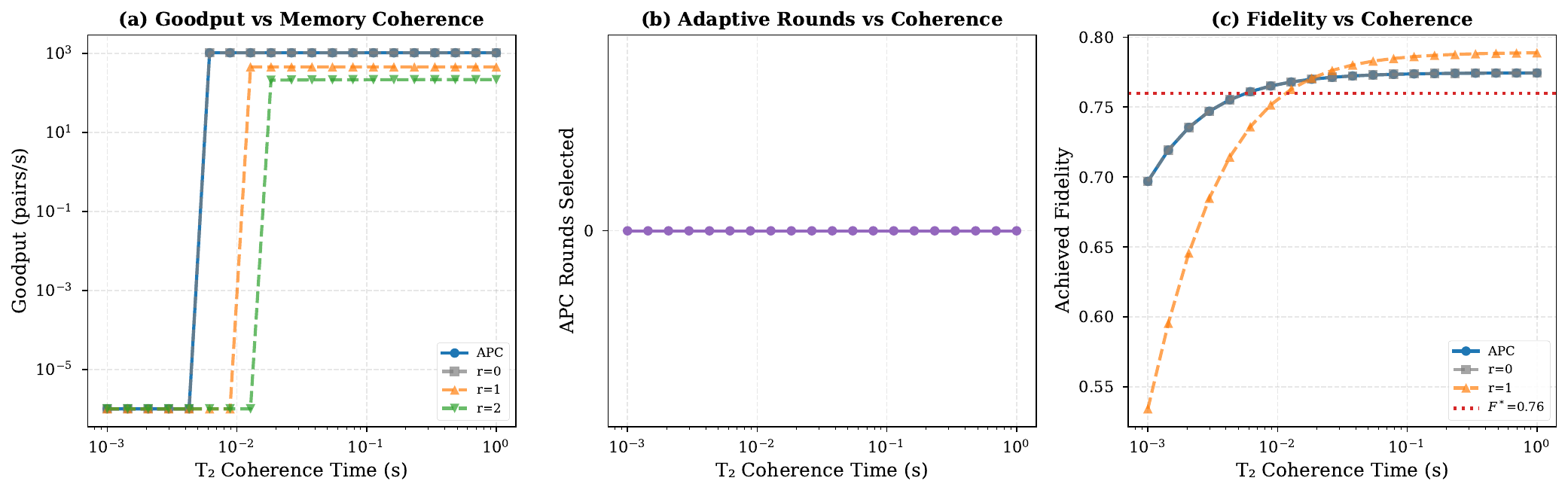}
  \caption{\textbf{Memory coherence time \(T_2\) induces a feasibility threshold.}
Setup: 3-hop chain over 24~km total, with per-link \(F_0=0.92\) calibrated so that the unpurified end-to-end fidelity is \(F_{\mathrm{raw}}\approx 0.77\) as \(T_2\rightarrow\infty\).
Target \(F^\star=0.76\); \(T_2\) swept from 1~ms to 1~s.
(a) Goodput versus \(T_2\) (log--log): below \(\sim 9\)~ms, decoherence prevents meeting \(F^\star\).
(b) APC-selected purification rounds versus \(T_2\): \(r=0\) throughout the viable region, indicating that once memory decay is sufficiently suppressed, swapping alone meets \(F^\star\).
(c) Achieved end-to-end fidelity versus \(T_2\), showing a sharp crossing at the feasibility threshold.}
  \label{fig:coherence}
\end{figure*}

\begin{figure*}[t!]
  \centering
  \includegraphics[width=\textwidth]{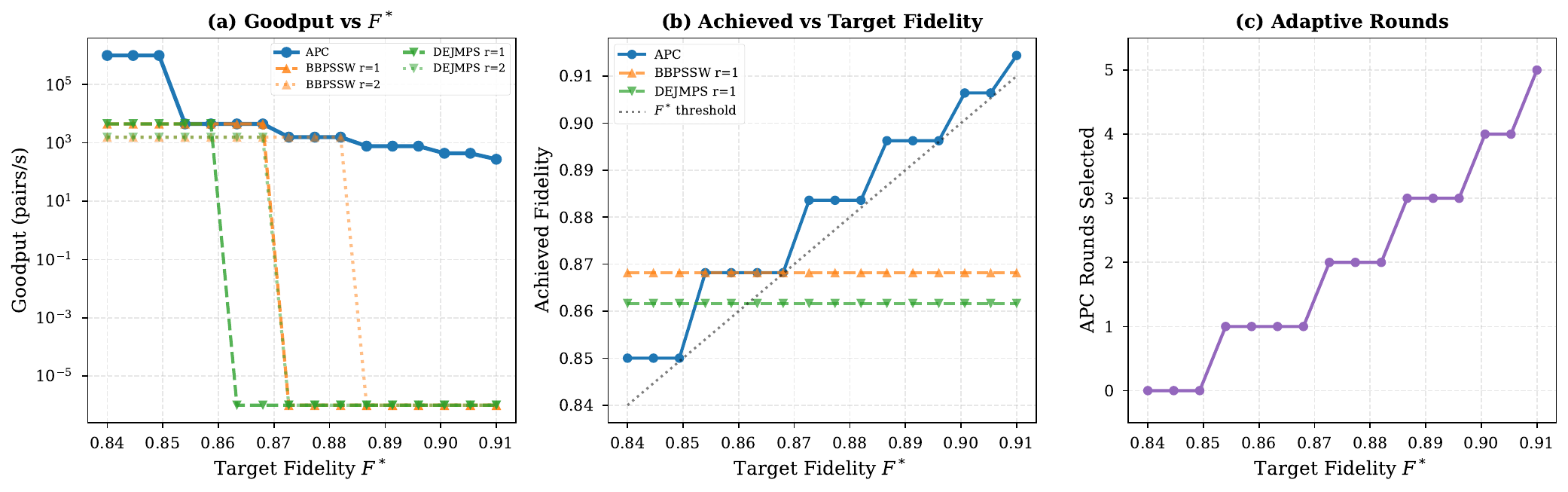}
  \caption{\textbf{Protocol/depth adaptation to maximize hard-threshold goodput.}
Setup: single hop with \(F_0=0.85\), distance 15~km, and \(T_2=150\)~ms.
We compare fixed BBPSSW and DEJMPS at \(r\in\{1,2\}\) and sweep \(F^\star\in[0.84,0.91]\).
(a) Goodput versus \(F^\star\) (semi-log): APC tracks the upper envelope by adapting depth (and protocol via the selection policy).
(b) Achieved fidelity versus target, showing APC meets the constraint while fixed baselines under- or over-shoot.
(c) APC-selected rounds versus \(F^\star\), exhibiting stepwise increases from \(r=0\) up to \(r=5\).}
  \label{fig:protocols}
\end{figure*}

\begin{figure*}[t!]
  \centering
  \includegraphics[width=\textwidth]{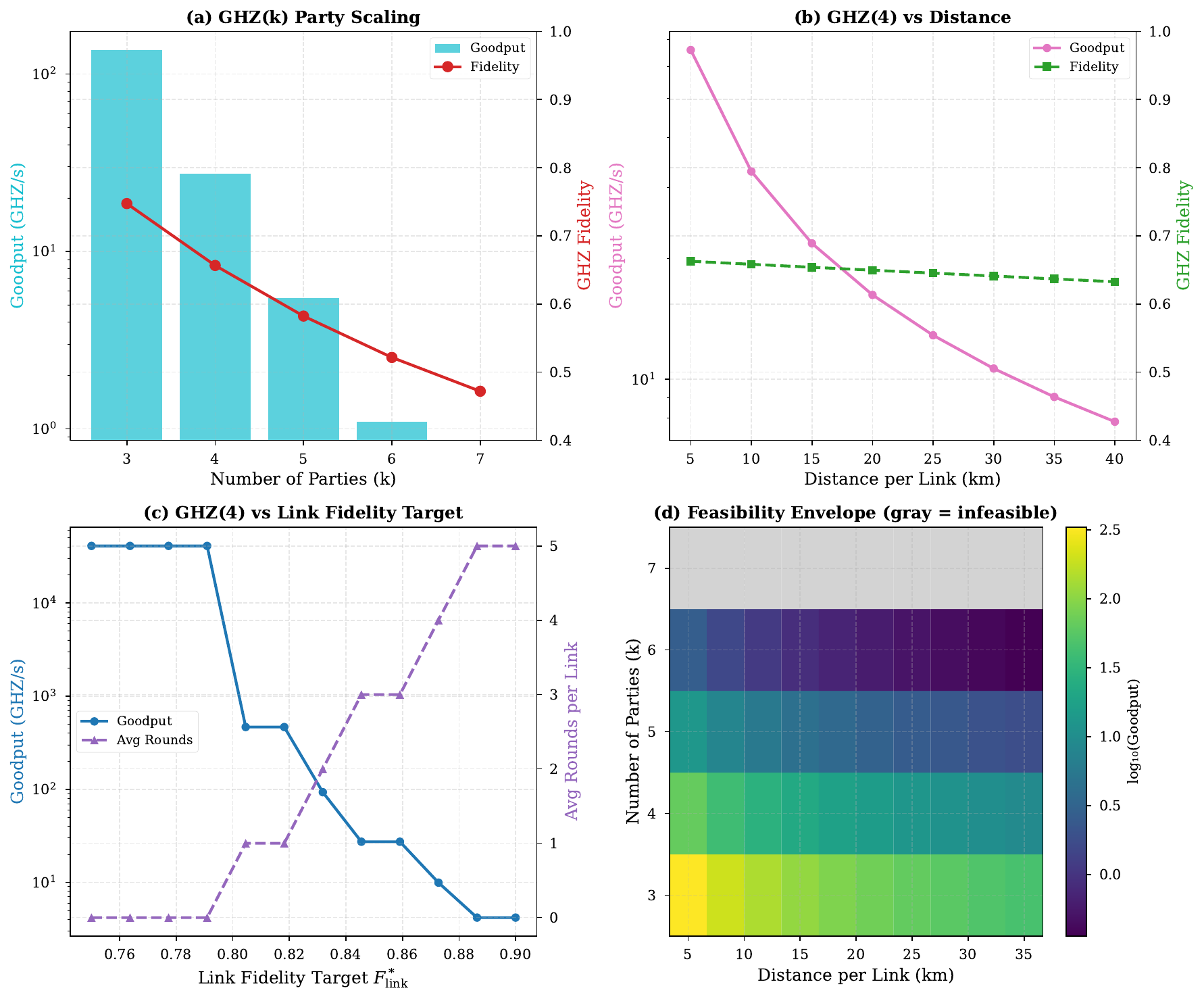}
  \caption{\textbf{Scaling and feasibility for \(\mathrm{GHZ}(k)\) distribution built from APC-optimized links.}
Setup: \(k\) parties connected by \(k-1\) links, per-link \(F_0=0.80\), per-link target \(F^\star_{\mathrm{link}}=0.85\), and 12~km per link unless otherwise stated.
(a) \(\mathrm{GHZ}(k)\) scaling: goodput (log scale) decreases rapidly with \(k\), consistent with multiplicative link success.
(b) \(\mathrm{GHZ}(4)\) versus distance per link: goodput decreases while GHZ fidelity remains comparatively stable over 5--40~km.
(c) \(\mathrm{GHZ}(4)\) versus \(F^\star_{\mathrm{link}}\): stricter link targets increase per-link rounds and reduce goodput.
(d) Feasibility heatmap over distance and \(k\) (gray=infeasible), showing that larger \(k\) requires shorter links to remain above threshold.}
  \label{fig:ghz}
\end{figure*}

\begin{figure*}[t!]
  \centering
  \includegraphics[width=\textwidth]{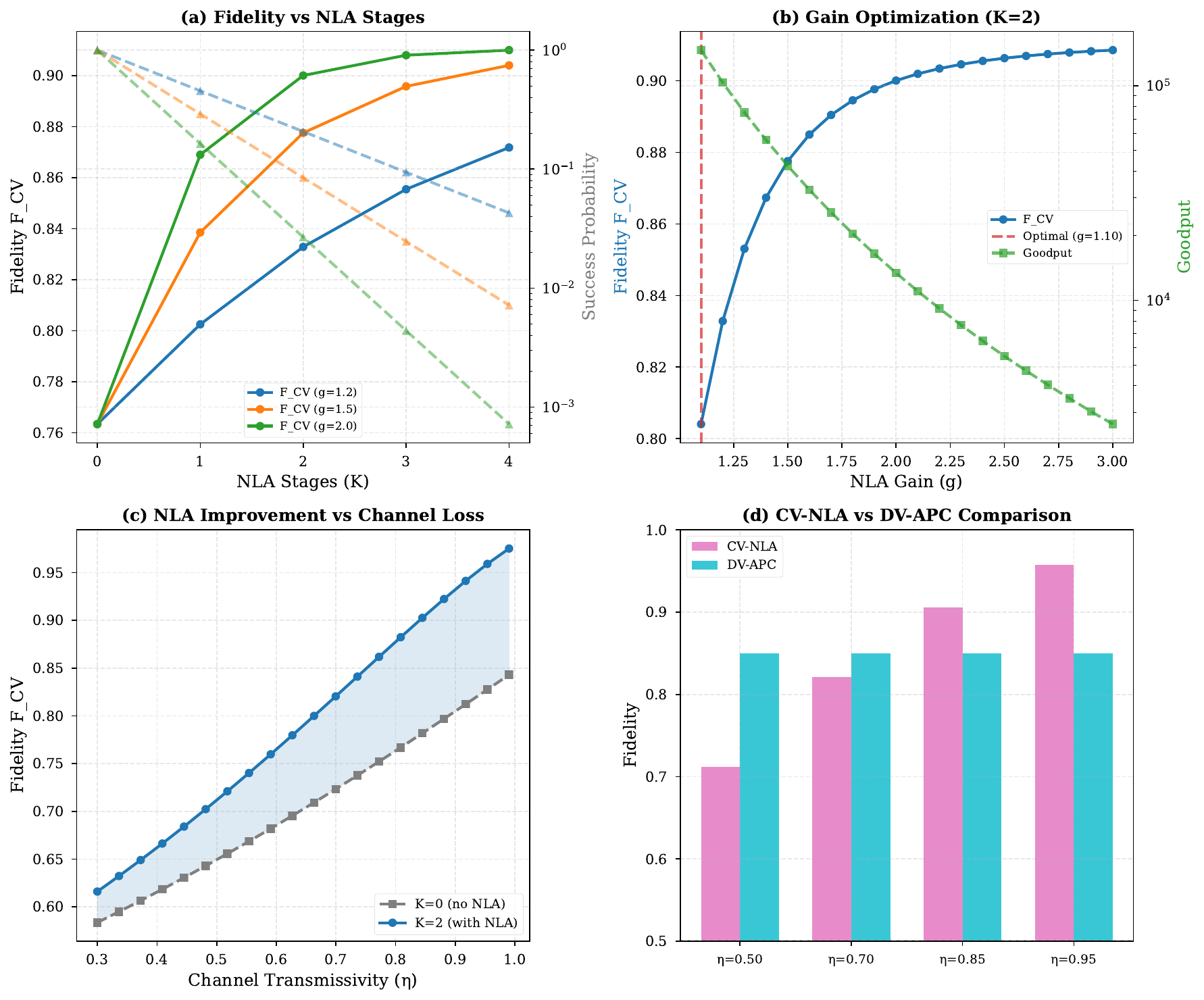}
  \caption{\textbf{CV NLA tradeoffs under APC's envelope model.} Setup: Gaussian CV states with initial squeezing $r=1.2$; channel transmissivity $\eta$ varied; NLA stages $K=0$ to $4$ with gain $g$. \textbf{(a)} $F_{\mathrm{CV}}$ versus $K$ for gains $g\in\{1.2,1.5,2.0\}$; higher gain improves $F_{\mathrm{CV}}$ but lowers success (right axis in dashed traces). \textbf{(b)} Gain optimization at $K=2$: $F_{\mathrm{CV}}$ rises with $g$ while goodput peaks near $g^\star\approx1.10$. \textbf{(c)} Improvement versus channel loss: $K=0$ versus $K=2$ across $\eta=0.3$ to $0.99$; the gap is largest at intermediate loss. \textbf{(d)} CV-NLA versus DV-APC at matched channel conditions $\eta\in\{0.50,0.70,0.85,0.95\}$; both fidelity and success are shown against the classical $F=0.5$ threshold.}
  \label{fig:cvnla}
\end{figure*}

\begin{figure*}[t!]
  \centering
  \includegraphics[width=\textwidth]{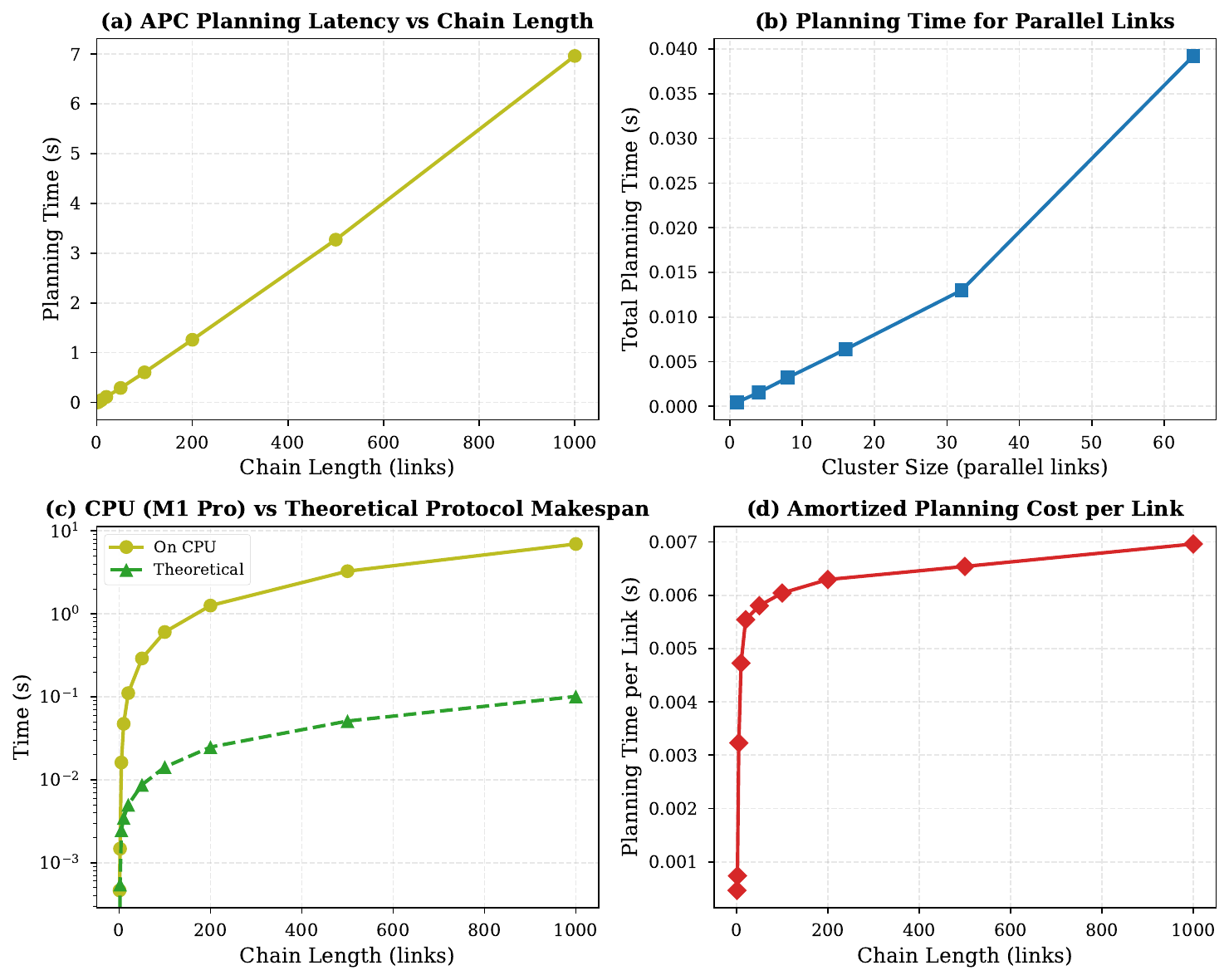}
  \caption{\textbf{APC planning is fast enough for real-time control.} \textbf{(a)} Planning time versus chain length for 1 to 1000 links (three calls averaged per point) with $F_0=0.88$, $p_{q}=0.9999$, $F^\star=0.85$, $T_2=1.0$ s; growth is roughly linear and remains near 6–7 s at 1000 links on an M1 Pro (Apple Silicon). \textbf{(b)} Total CPU time to plan $m$ independent single-link requests for $m\in\{1,4,8,16,32,64\}$; sequential planning scales linearly, indicating parallelization headroom. \textbf{(c)} CPU planning time versus theoretical protocol makespan (log scale) across chain length; planning is negligible for small chains and remains small compared to execution even at large scale. \textbf{(d)} Amortized planning time per link decreases with chain length, approaching a few milliseconds per link around 1000 links.}
  \label{fig:timing}
\end{figure*}
\mbox{}

\subsubsection{Path composition via frontier-based dynamic programming}
Rather than discretizing fidelity on a fixed grid, the APC planner propagates a \emph{compact set of candidate partial plans} (a Pareto frontier) along the path.
Each candidate carries the sufficient statistics needed for objective evaluation and constraint checking:
\begin{equation}
x_j \;=\; \big(F_j,\; C_j,\; T^{\mathrm{gen}}_j,\; T^{\mathrm{swap}}_j,\; P_j,\; \Pi_j \big),
\end{equation}
where \(F_j\) is the prefix fidelity after composing the first \(j\) links and the corresponding swaps,
\(C_j\) is expected raw EPR consumption,
\(T^{\mathrm{gen}}_j\) and \(T^{\mathrm{swap}}_j\) are expected generation and swapping time components,
\(P_j\) is the end-to-end success probability factor accumulated so far, and
\(\Pi_j\) records the sequence of per-link decisions.

\paragraph{One-step extension.}
Given a prefix candidate \(x_{j-1}\) and a new link-level choice \(u_j\) (purification method and depth on link \(j\)),
the physics backend provides \((F^{\mathrm{link}}_j,\, p^{\mathrm{pur}}_j,\, C^{\mathrm{link}}_j,\, t^{\mathrm{pur}}_j)\).
We then update:
\begin{align}
F_j &= \mathrm{SwapCompose}\!\left(F_{j-1}, F^{\mathrm{link}}_j\right), \\
P_j &= P_{j-1}\cdot p^{\mathrm{pur}}_j \cdot p^{\mathrm{BSM}}_j, \\
C_j &= C_{j-1} + C^{\mathrm{link}}_j, \\
T^{\mathrm{gen}}_j &= \mathrm{GenAgg}\!\left(T^{\mathrm{gen}}_{j-1},\; t^{\mathrm{gen}}_j\right), \\
T^{\mathrm{swap}}_j &= T^{\mathrm{swap}}_{j-1} + t^{\mathrm{swap}}_j,
\end{align}
where \(\mathrm{GenAgg}(\cdot)\) is either a parallel aggregation (e.g., \(\max\)) or sequential aggregation (sum), and
\(p^{\mathrm{BSM}}_j, t^{\mathrm{swap}}_j\) capture swapping success and time at the intermediate node.

\paragraph{Pareto pruning.}
After generating all extensions for prefix length \(j\), we prune dominated candidates.
A candidate \(x\) dominates \(x'\) if it is no worse in all of \((F,\;T^{\mathrm{gen}}{+}T^{\mathrm{swap}},\;C)\) and strictly better in at least one.
To keep planning latency bounded, we retain at most a fixed number of non-dominated candidates (the frontier width) per prefix.
This yields a fast approximate planner whose accuracy improves monotonically as the frontier width is increased.

\paragraph{Plan selection.}
After processing all hops, we evaluate the remaining candidates using the chosen objective (e.g., hard-threshold goodput)
and return the best feasible plan (or the closest-to-feasible plan when the constraint \(F_{\mathrm{end}}\ge F^\star\) is unattainable).

\subsubsection{Objectives and Goodput}
APC currently supports several scalar objectives:
\begin{itemize}
  \item minimizing time, then minimizing EPR cost:
    \begin{equation}
      J(\mathbf{u}) = \big(T_{\text{makespan}}(\mathbf{u}), C_{\text{pairs}}^{\text{path}}(\mathbf{u})\big),
    \end{equation}
    with lexicographic minimization;
  \item minimizing EPR cost, then minimizing time:
    \begin{equation}
      J(\mathbf{u}) = \big(C_{\text{pairs}}^{\text{path}}(\mathbf{u}), T_{\text{makespan}}(\mathbf{u})\big);
    \end{equation}
  \item Pareto selection over time and cost, exposing the frontier instead of a single point;
  \item goodput maximization, defined as
    \begin{equation}
      G(\mathbf{u}) =
      \begin{cases}
        0, & F_{\text{end}}(\mathbf{u}) < F^\star,\\[0.3em]
        \dfrac{P_{\text{succ}}^{\text{path}}(\mathbf{u})}{T_{\text{makespan}}(\mathbf{u})}, & F_{\text{end}}(\mathbf{u}) \ge F^\star,
      \end{cases}
    \end{equation}
    which is implemented by minimizing $J(\mathbf{u}) = -G(\mathbf{u})$.
\end{itemize}

Because static purification schemes that apply the same protocol and depth on every hop correspond to restricting $\mathcal{P}_{\ell_1} = \cdots = \mathcal{P}_{\ell_H} = \{u_{\text{static}}\}$, they are represented as a subset of APC's design space. For any additive or monotone objective $J$, the global optimum over APC's full design space is therefore never worse than the optimum over static schemes; this is illustrated empirically in Section~\ref{section:results}.

\section{Experimental Setup}
\label{sec:exp-setup}

To evaluate the efficacy of the APC in realistic 6G quantum networking scenarios, we conducted a comprehensive suite of benchmarks. The experiments are designed to isolate the impact of specific physical parameters like memory coherence, gate error rates, and topology size on the achievable application utility, quantified here as goodput (valid entangled pairs per second).

In addition to per-link purification before swapping, the implementation optionally supports a final end-to-end recurrence stage applied after swapping between the end nodes (disabled in the default benchmarks unless stated).
This allows trading additional end-to-end attempts for higher delivered fidelity when intermediate memory and gate noise permit.

\subsection{Simulation Environment}

Experiments were performed using the KOSMOS discrete-event simulator with a high-performance fidelity-based physics backend. Unlike state-vector or density-matrix simulations whose cost grows exponentially with qubit number, our fidelity-level backend uses closed-form analytic maps for Werner and Bell-diagonal inputs (BBPSSW/DEJMPS~\cite{Bennett1996Purification,Deutsch1996QPA}). This enables rapid evaluation of deep recurrence and long repeater chains (hundreds of links). The timing layer is latency-aware: it includes classical round-trip times and models memory aging during waits (via an exact expectation for parallel link generation and $T_2$ dephasing), so makespan and goodput reflect scheduling and storage effects. 

\subsection{Network Topologies and Physics Model}

We focus on two canonical topologies relevant to 6G quantum networks:

\begin{enumerate}
  \item Linear Repeater Chain: A multi-hop path connecting a source and destination via homogeneous repeater nodes. Unless otherwise stated, links are modeled as optical fibers with an attenuation of $0.2$ dB/km and a signal speed of $2\times 10^8$ m/s.

  \item Multipartite Star: A central hub distributing entanglement to $k$ leaf nodes to form GHZ states, representative of a quantum sensor network or distributed computing cluster.
\end{enumerate}

\section{Results and Discussion}
\label{section:results}

\subsection{Maximizing Goodput via Adaptive Depth}

The core advantage of the APC is its ability to navigate the trade-off between fidelity and latency dynamically. Fig.~\ref{fig:goodput} illustrates this behavior for a single 15 km link. As the target fidelity $F^\star$ increases, static strategies exhibit a step-like failure mode: $r=0$ fails immediately when $F_{\text{raw}} < F^\star$, while $r=2$ produces pairs with $F \gg F^\star$ but at a significantly reduced rate due to the probabilistic nature of recurrence.
In contrast, the APC traces the Pareto-optimal envelope of these static curves (Fig.~\ref{fig:goodput}a). By automatically switching from $r=0$ to $r=1$ and $r=2$ only when necessary (Fig.~\ref{fig:goodput}b), the controller maintains the highest possible goodput. Notably, the APC minimizes ``fidelity overshoot'' (Fig.~\ref{fig:goodput}c), ensuring that network resources are not squandered on generating hyper-pure states when the application request does not demand them.

\subsection{Feasibility Envelope and Noise Robustness}

The operational limits of the network are defined by the ``feasibility envelope'', the region in parameter space where $F_{\text{end}} \ge F^\star$ is physically achievable. Fig.~\ref{fig:feasible} delineates this boundary across link distance and target fidelity. We observe a sharp cutoff where the combination of transmission loss (lowering rate) and classical signaling latency (increasing decoherence) renders high-fidelity distribution impossible, regardless of the purification depth.
This limitation is further dissected in Fig.~\ref{fig:noise}, which isolates the impact of gate errors. While purification can suppress channel noise, it accumulates local gate errors. Here, we use a correlated noise sweep $p_1=p_2=p_{\mathrm{meas}}=\varepsilon$ to reduce parameter space (results should be interpreted as joint scaling of local error sources). Consequently, we observe a ``noise cliff'' at gate error rates $\approx 10^{-3}$. Beyond this threshold, aggressive purification ($r \ge 2$) becomes detrimental, as the entropy added by noisy CNOT gates outweighs the distillation gain. The APC correctly identifies this regime (Fig.~\ref{fig:noise}c), truncating the purification depth to maximize utility in high-noise environments.

\subsection{Impact of Memory Coherence}

A critical constraint in wide-area quantum networks is memory coherence time ($T_2$). As shown in Fig.~\ref{fig:coherence}, there exists a minimum $T_2$ required to meet the fidelity constraint under the end-to-end timing model (and, if needed, enable recurrence). For a 3-hop chain targeting $F^\star=0.76$, coherence times below $10$ ms render the protocol infeasible (Fig.~\ref{fig:coherence}a). This is because the classical round-trip time required to herald purification outcomes allows stored qubits to decohere below the distillation threshold. The APC adapts to this by disabling purification ($r=0$) in low-$T_2$ regimes (Fig.~\ref{fig:coherence}b), effectively reverting to a ``repeat-until-success'' swapping strategy to salvage connectivity, albeit at lower fidelity.

\subsection{When protocol identity matters}

Fig.~\ref{fig:protocols}--Under Werner noise, BBPSSW and DEJMPS provide nearly identical fidelity gains per round. The APC's advantage therefore comes from choosing the right number of rounds, not the protocol identity. As $F^\star$ increases, the APC climbs in $r$ while single-protocol static strategies either waste effort by overshooting or fail the constraint. 

Beyond adjusting recurrence depth, the APC can \emph{select between purification protocol families} (BBPSSW vs.\ DEJMPS) through a lightweight protocol-selection policy based on the assumed noise structure and hardware regime.
In the current implementation, this policy biases toward BBPSSW for (approximately) Werner-like inputs and toward DEJMPS when asymmetric Bell-diagonal structure or strong two-qubit gate noise makes DEJMPS preferable.
Fig.~\ref{fig:protocols} illustrates that, in an asymmetric regime, DEJMPS enables stricter \(F^\star\) targets that BBPSSW baselines fail to reach, while the APC adapts purification depth to remain on the goodput envelope.

\subsection{Multipartite and Continuous Variable Extensions}

We extended the evaluation to multipartite GHZ state distribution (Fig.~\ref{fig:ghz}) and continuous variable entanglement (Fig.~\ref{fig:cvnla}).
For GHZ states, the goodput scales as $\sim p_{\text{link}}^{k}$ (the per-link success probability after purification), resulting in an exponential decay with the number of parties $k$ (Fig.~\ref{fig:ghz}a). The APC mitigates this by optimizing the underlying bipartite links, but the results highlight the necessity of high-efficiency fusion gates for scaling beyond $k=5$.
For CV systems, we simulated Noiseless Linear Amplification (NLA). Fig.~\ref{fig:cvnla} confirms that optimal gain choices ($g \approx 1.1-1.5$) can counteract channel loss, though the probabilistic nature of NLA imposes a heavy rate penalty similar to discrete variable recurrence.

\subsection{Real-Time Planning Latency}

Finally, we benchmarked the computational overhead of the APC planner itself (Fig.~\ref{fig:timing}). The dynamic programming algorithm exhibits near-linear scaling with chain length. For a 1000-link chain, the planning latency is approximately 6--7 seconds on a M1 Pro processor, with a per-link amortized cost of milliseconds. This confirms that the APC is computationally lightweight enough to operate in the control plane of a software-defined quantum network, calculating optimal policies on timescales faster than the drift of physical hardware parameters.

\section{Conclusion}
\label{sec:conclusion}

In this work, we presented the Adaptive Purification Controller (APC), a robust control plane module for maximizing utility in 6G quantum networks. By formulating entanglement distribution as a resource optimization problem constrained by fidelity, the APC demonstrated significant advantages over static protocols:

\begin{enumerate}
  \item Optimal Resource Usage: It consistently identifies the minimum recurrence depth required to meet application needs, maximizing goodput and freeing resources for other requests.

  \item Hardware Agnosticism: The controller adapts seamlessly to varying device parameters, identifying ``noise cliffs'' and coherence thresholds where purification becomes detrimental.

  \item Scalability: Benchmarks confirm the planner's feasibility for large-scale networks, supporting both discrete-variable repeater chains and multipartite sensor networks.

  \item Low Latency: The millisecond-scale planning latency (Fig.~\ref{fig:timing}) confirms that APC is lightweight enough to be embedded in the Software-Defined Networking (SDN) control plane of future 6G hybrid networks, enabling real-time orchestration of classical and quantum flows.
\end{enumerate}

Future work will integrate the APC with distributed quantum error correction focusing on erasure-aware decoding, extending its outputs to erasure-aware logical link budgeting and guardrailing for fault-tolerant nonlocal operations in modular networks. This will further support quantum error correction codes for third-generation repeater networks. The APC thus represents a key step toward the automated, utility-driven management of the Quantum Internet.

\bibliographystyle{IEEEtran}\bibliography{bibliography}

@article{Bennett1996Purification,
  title = {Purification of Noisy Entanglement and Faithful Teleportation via Noisy Channels},
  author = {Bennett, Charles H. and Brassard, Gilles and Popescu, Sandu and Schumacher, Benjamin and Smolin, John A. and Wootters, William K.},
  journal = {Phys. Rev. Lett.},
  volume = {76},
  issue = {5},
  pages = {722--725},
  numpages = {0},
  year = {1996},
  month = {Jan},
  publisher = {American Physical Society},
  doi = {10.1103/PhysRevLett.76.722}
}

@article{Deutsch1996QPA,
  title = {Quantum Privacy Amplification and the Security of Quantum Cryptography over Noisy Channels},
  author = {Deutsch, David and Ekert, Artur and Jozsa, Richard and Macchiavello, Chiara and Popescu, Sandu and Sanpera, Anna},
  journal = {Phys. Rev. Lett.},
  volume = {77},
  issue = {13},
  pages = {2818--2821},
  numpages = {0},
  year = {1996},
  month = {Sep},
  publisher = {American Physical Society},
  doi = {10.1103/PhysRevLett.77.2818}
}

@article{DevetakWinter2005,
  author={Devetak, I. and Winter, A.},
  title={Distillation of secret key and entanglement from quantum states},
  journal={Proc. R. Soc. A},
  volume={461},
  number={2053},
  pages={207--235},
  month={Jan},
  year={2005},
  doi={10.1098/rspa.2004.1379}
}

@article{Dur2003GraphStates,
  title = {Multiparticle Entanglement Purification for Graph States},
  author = {D\"ur, W. and Aschauer, H. and Briegel, H.-J.},
  journal = {Phys. Rev. Lett.},
  volume = {91},
  issue = {10},
  pages = {107903},
  numpages = {4},
  year = {2003},
  month = {Sep},
  publisher = {American Physical Society},
  doi = {10.1103/PhysRevLett.91.107903}
}

@article{Aschauer2005TwoColorable,
  title = {Multiparticle entanglement purification for two-colorable graph states},
  author = {Aschauer, H. and D\"ur, W. and Briegel, H.-J.},
  journal = {Phys. Rev. A},
  volume = {71},
  issue = {1},
  pages = {012319},
  numpages = {20},
  year = {2005},
  month = {Jan},
  publisher = {American Physical Society},
  doi = {10.1103/PhysRevA.71.012319}
}

@ARTICLE{deBone2020GHZ,
  author={de Bone, Sebastian and Ouyang, Runsheng and Goodenough, Kenneth and Elkouss, David},
  journal={IEEE Transactions on Quantum Engineering}, 
  title={Protocols for Creating and Distilling Multipartite GHZ States With Bell Pairs}, 
  year={2020},
  volume={1},
  number={},
  pages={1-10},
  doi={10.1109/TQE.2020.3044179}}

@article{Eisert2004CV,
title = {Distillation of continuous-variable entanglement with optical means},
author = {J. Eisert and D.E. Browne and S. Scheel and M.B. Plenio},
journal = {Annals of Physics},
volume = {311},
number = {2},
pages = {431-458},
year = {2004},
issn = {0003-4916},
doi = {https://doi.org/10.1016/j.aop.2003.12.008},
}

@article{Fiurasek2009Gaussian,
  author = {Fiurášek, Jaroslav},
  title = {Gaussian transformations and distillation of entangled Gaussian states},
  journal = {Physical Review A},
  volume = {80},
  number = {5},
  pages = {053822},
  year = {2009},
  doi = {10.1103/PhysRevA.80.053822},
  publisher = {American Physical Society}
}

@article{Xiang2010NLA,
  author={Xiang G., Ralph T., Lund A. et al.},
  title={Heralded noiseless linear amplification and distillation of entanglement},
  journal={Nature Photon},
  volume={4},
  pages={316--319},
  month={May},
  year={2010},
  doi={https://doi.org/10.1038/nphoton.2010.35}
}

@article{Guanzon2024NLA,
  title = {Saturating the Maximum Success Probability Bound for Noiseless Linear Amplification Using Linear Optics},
  author = {Guanzon, Joshua J. and Winnel, Matthew S. and Singh, Deepesh and Lund, Austin P. and Ralph, Timothy C.},
  journal = {PRX Quantum},
  volume = {5},
  issue = {2},
  pages = {020359},
  numpages = {14},
  year = {2024},
  month = {Jun},
  publisher = {American Physical Society},
  doi = {10.1103/PRXQuantum.5.020359}
}

@article{Zukowski1993Swapping,
  title = {``Event-ready-detectors'' Bell experiment via entanglement swapping},
  author = {\ifmmode \dot{Z}\else \.{Z}\fi{}ukowski, M. and Zeilinger, A. and Horne, M. A. and Ekert, A. K.},
  journal = {Phys. Rev. Lett.},
  volume = {71},
  issue = {26},
  pages = {4287--4290},
  numpages = {0},
  year = {1993},
  month = {Dec},
  publisher = {American Physical Society},
  doi = {10.1103/PhysRevLett.71.4287}
}

@article{Shchukin2017Waiting,
  title = {Waiting time in quantum repeaters with probabilistic entanglement swapping},
  author = {Shchukin, E. and Schmidt, F. and van Loock, P.},
  journal = {Phys. Rev. A},
  volume = {100},
  issue = {3},
  pages = {032322},
  numpages = {20},
  year = {2019},
  month = {Sep},
  publisher = {American Physical Society},
  doi = {10.1103/PhysRevA.100.032322}
}

@article{NetSquid2021,
  author = {Coopmans T., Knegjens R., Dahlberg A. et al.},
  title = {NetSquid, a NETwork Simulator for QUantum Information using Discrete events},
  journal = {Communications Physics},
  volume = {4},
  number = {1},
  pages = {164},
  year = {2021},
  doi = {10.1038/s42005-021-00647-8},
  publisher = {Nature Publishing Group}
}

@article{NetSquidFidelity2024,
AUTHOR = {Pérez Castro, David and Fernández Vilas, Ana and Fernández Veiga, Manuel and Blanco Rodríguez, Mateo and Díaz Redondo, Rebeca P.},
TITLE = {Simulation of Fidelity in Entanglement-Based Networks with Repeater Chains},
JOURNAL = {Applied Sciences},
VOLUME = {14},
YEAR = {2024},
NUMBER = {23},
ARTICLE-NUMBER = {11270},
ISSN = {2076-3417},
DOI = {10.3390/app142311270}
}

@article{NetSquidEBN2025,
  title={Simulation of entanglement based quantum networks for performance characterization},
  author={David P{\'e}rez-Castro and Juan Fern'andez-Herrer'in and Ana Fern{\'a}ndez Vilas and Manuel Fern{\'a}ndez-Veiga and Rebeca P. D{\'i}az Redondo},
  journal={ArXiv},
  year={2025},
  volume={abs/2501.03210},
}

@article{SimQN2023,
  author={Chen, Lutong and Xue, Kaiping and Li, Jian and Yu, Nenghai and Li, Ruidong and Sun, Qibin and Lu, Jun},
  journal={IEEE Network}, 
  title={SimQN: A Network-Layer Simulator for the Quantum Network Investigation}, 
  year={2023},
  volume={37},
  number={5},
  pages={182-189},
  doi={10.1109/MNET.130.2200481}}

@misc{Sequence,
  author={Behera, B. K. and Das, A. and Nandi, A. and Sen(De), A. and Sen, U.},
  title={SeQUeNCe: A Simulator of QUantum Network Communication},
  howpublished={\url{https://github.com/sequence-toolbox/SeQUeNCe}},
  year={2020}
}

@article{QuNetSim,
  title={QuNetSim: A Software Framework for Quantum Networks},
  author={DiAdamo, Stephen and N{\"o}tzel, Janis and Zanger, Benjamin and Be{\c{s}}e, Mehmet Mert},
  journal={IEEE Transactions on Quantum Engineering},
  year={2021},
  doi={10.1109/TQE.2021.3092395}
}

@misc{shi2024designentanglementpurificationprotocol,
      title={Design of an entanglement purification protocol selection module}, 
      author={Yue Shi and Chenxu Liu and Samuel Stein and Meng Wang and Muqing Zheng and Ang Li},
      year={2024},
      eprint={2405.02555},
      archivePrefix={arXiv},
      primaryClass={quant-ph}
}

@article{2023Purification,
  title = {Entanglement purification on quantum networks},
  author = {Victora, Michelle and Tserkis, Spyros and Krastanov, Stefan and de la Cerda, Alexander Sanchez and Willis, Steven and Narang, Prineha},
  journal = {Phys. Rev. Res.},
  volume = {5},
  issue = {3},
  pages = {033171},
  numpages = {11},
  year = {2023},
  month = {Sep},
  publisher = {American Physical Society},
  doi = {10.1103/PhysRevResearch.5.033171}
}

\end{document}